\documentclass[a4paper,12pt]{article}

\usepackage{theorem}
\usepackage{amsfonts}

\theoremstyle{change}

\theorembodyfont{\rmfamily}

\def\Box{\hbox{\raisebox{0.0em}{\rlap{$\sqcap$}}\kern0em%
            \raisebox{-0.0em}{$\sqcup$}} } 
\newenvironment{proof}{{\it Proof. }}{\hfill$\Box$\vspace{0.5cm}}
\def\bepf{\begin{proof}}
\def\epf{\end{proof}}

\def\fct#1{\mathop{\rm #1}}	

\def\div{\fct{div}}

\def\im{\fct{Im}}

\def\re{\fct{Re}}

\def\hbar{h\hspace{-2mm}^-}

\def\phi{\varphi}

\def\wave{\protect{\footnotesize $\sim$}}

\def\<{\langle} 				
\def\>{\rangle} 				
\def\rangu{\hbox{\raisebox{0.15em}{\rlap{$\sqcap$}}\kern0em%
            \raisebox{-0.19em}{$\sqcup$}} } 

\def\beq{\begin{equation}} 
\def\eeq{\end{equation}} 
\def\lbeq#1{\begin{equation} \label{#1}} 
\def\beqar{\begin{eqnarray}}
\def\eeqar{\end{eqnarray}}
\def\bary{\begin{array}}
\def\eary{\end{array}}
\def\becas{\left\{ \begin{array}{l@{\qquad}l}}
\def\ecas{\end{array} \right.}
\def\benu{\begin{enumerate}}
\def\eenu{\end{enumerate}}
\def\gzit#1{{\rm (\ref{#1})}} 			

\begin{document}

\vspace*{-2cm}

\begin{center}

{\LARGE \bf Bohmian mechanics contradicts quantum mechanics} \\

\vspace{1cm}

\centerline{\sl {\large \bf Arnold Neumaier}}

\vspace{0.5cm}

\centerline{\sl Institut f\"ur Mathematik, Universit\"at Wien}
\centerline{\sl Strudlhofgasse 4, A-1090 Wien, Austria}
\centerline{\sl email: neum@cma.univie.ac.at}
\centerline{\sl WWW: http://solon.cma.univie.ac.at/\wave neum/}

\end{center}

\vspace{0.5cm}

{\bf Abstract.} 

It is shown that, for a harmonic oscillator in the ground state, 
Bohmian mechanics and quantum mechanics predict values of opposite 
sign for certain time correlations. 

The discrepancy can be explained by the fact that Bohmian mechanics
has no natural way to accomodate the Heisenberg picture, since the
local expectation values that define the beables of the theory depend
on the Heisenberg time being used to define the operators.

Relations to measurement are discussed, too, and shown to leave no
loophole for claiming that Bohmian mechanics reproduces {\em all}
predictions of quantum mechanics {\em exactly}.

\vfill

\begin{flushleft}

\vspace{0.5cm}

{\bf 1998\hspace{.4em} PACS Classification}: 
             03.65.Bz\\

{\bf E-print Archive No.}: quant-ph/0001011

\end{flushleft}

\newpage 
\section{Introduction} \label{In}

Since its inception by {\sc Bohm} \cite{Boh} and its popularization
by {\sc Bell} \cite{Bel}, the pilot wave theory, or causal 
interpretation of quantum mechanics -- now often called Bohmian 
mechanics -- has been regarded by a number of people as a in some 
respects bizarre but otherwise viable ontology for quantum mechanics. 
Books and proceedings appeared that discuss the
features of the theory in detail, cf. {\sc Holland} \cite{Hol},
{\sc Bohm \& Hiley} \cite{BohH}, {\sc Cushing} et al. \cite{CusFG},
good introductory surveys are available,
cf. {\sc Berndl} et al. \cite{BerDD}, {\sc D\"urr} et al. \cite{DueGZ},
and accounts for the lay reader exist, cf. {\sc Albert} \cite{Alb},
{\sc Goldstein} \cite{Gol}. 

On the other hand, Bohmian mechanics has remained a minority view, 
since, from its beginnings, it had been critically viewed by most of 
the influential quantum physicists. The main early arguments against
it are stated in {\sc Holland} \cite[Sections 1.5.3 and 6.5.3]{Hol}; 
they are usually argued away by some mathematical analysis accompanied 
by statements such as ``classical prejudice''
({\sc Bell} \cite[Chapter 14]{Bel}),``to our knowledge no serious 
technical objections have ever been raised against'' it ({\sc Holland} 
\cite[Section 1.5.3]{Hol}), or ``Bohmian mechanics accounts for all of
the phenomena governed by nonrelativistic quantum mechanics'' 
({\sc D\"urr} et al. \cite{DueGZ}). The arguments on both sides 
usually rest on one's unwillingness or readiness to accept 
counterintuitive consequences of the Bohmian picture, since none of 
the phenomena in question are observable.

More recent counterintuitive implications of Bohmian mechanics 
({\sc Englert} et al. \cite{EngSS}, {\sc Griffiths} \cite{Gri}) met 
with similar responses ({\sc D\"urr} et al. \cite{DueFG}, 
{\sc Dewdney} et al. \cite{DewHS}). In particular, D\"urr et al. write,
``an open-minded advocate of quantum orthodoxy would presumably have 
preferred the clearer and stronger claim that BM is {\it incompatible} 
with the predictions of quantum theory, so that, despite its virtues, 
it would not in fact provide an explanation of quantum phenomena. The 
authors are, however, aware that such a strong claim would be false.''

The purpose of this paper is to demonstrate -- independent of the
arguments in \cite{EngSS,Gri,Hol} -- that such a strong claim 
is valid indeed. Specifically, Bohmian mechanics contradicts the 
predictions of quantum mechanics at the level of time correlations. 
Since time correlations can be observed experimentally via linear 
response theory (see, e.g., {\sc Reichl} \cite[Chapter 15.H]{Rei}), 
Bohmian mechanics and quantum mechanics cannot be both valid.

Concerning discrepancies between Bohmian mechanics and 
quantum mechanics involving multiple times, see also 
{\sc Redington} et al. \cite{RedWS} for Bohmian hydrogen atoms, and 
{\sc Ghose} \cite{Gho} for histories of indistinguishable particles.

There are similar problems with multiple times in {\sc Nelson}'s 
\cite{Nel} stochastic quantum mechanics; however, there they can be 
overcome by a specific procedure for state reduction under 
measurement, see {\sc Blanchard} et al. \cite{BlaGS}. 
Bohmian mechanics does not seem to have such an option to rescue their 
case since in the orthodox Bohmian interpretation state 
reduction is a purely dynamical phenomenon.

\bigskip
{\bf Acknowledgments.}
I'd like to thank Philippe Blanchard, Sheldon Goldstein, 
Arkadiusz Jadzcyk and Jack Sarfatti for their comments on an earlier 
version of this paper.

\bigskip
\section{Background} 

{\bf Quantum mechanics.}
A one-dimensional quantum particle without spin in an external 
potential $V(q)$ is described by the Hamiltonian 
\lbeq{e.ham}
H(p,q)=\frac{p^2}{2m}+V(q)
\eeq
(see, e.g., {\sc Messiah} \cite[(2.20)]{Mes1}),
where the position operator $q$ and the momentum operator $p$
satisfy the canonical commutation relations 
\lbeq{e.ccr}
[q,p]=i\hbar
\eeq
\cite[(5.53)]{Mes1}.
In the Schr\"odinger picture, observables are associated with
Hermitian operators $A$.
The dynamics of a quantity $A$ is given in the Heisenberg picture by 
one-parameter families of operators $A(t)$ satisfying 
\lbeq{e.heis}
i\hbar \dot A(t)=[A(t),H(p(t),q(t))]
\eeq
\cite[(8.40)]{Mes1}; the identification with the Schr\"odinger
picture is obtained by specifying the initial condition $A(0)=A$
at some reference time $t=0$.

In the position representation, pure ensemble states are given by   
wave functions $\psi_0(x)$ satisfying $\int |\psi_0(x)|^2dx=1$, on 
which $q$ acts as multiplication by $x$ and $p$ acts as the differential
operator $\frac{\hbar}{i}\nabla$. The expectation of a Heisenberg 
operator family $A(t)$ in a pure ensemble is defined by
\lbeq{e.expQ}
\<A(t)\>_Q=\int \psi_0^*(x)(A(t)\psi_0)(x)dx
\eeq
\cite[(4.22)]{Mes1}.
If one defines a time-dependent wave function $\psi(x,t)$ as the
solution of the initial-value problem 
\lbeq{e.schr}
i\hbar\frac{\partial}{\partial t}\psi(x,t)=H\psi(x,t),
~~~\psi(x,0)=\psi_0(x)
\eeq
\cite[(2.29)]{Mes1},
one can rewrite the expectation in the equivalent Schr\"odinger picture 
as 
\lbeq{e.expQ0}
\<A(t)\>_Q=\int \psi^*(x,t)(A\psi)(x,t)dx
\eeq
\cite[(4.22)]{Mes1}.
In particular, the expectation of a function of position is
\lbeq{e.exppos}
\<f(q(t))\>_Q=\int f(x)|\psi(x,t)|^2dx
\eeq
\cite[(4.13)]{Mes1},
so that 
\lbeq{e.prob}
P(x,t)=|\psi(x,t)|^2
\eeq
\cite[(4.2)]{Mes1}
behaves as a probability density. For Hamiltonians of the form 
\gzit{e.ham}, the probability density satisfies an equation of 
continuity,
\lbeq{e.cont}
\frac{\partial}{\partial t}P+\div J=0
\eeq
\cite[(4.11)]{Mes1},
with the probability current 
\lbeq{e.curr}
J(x,t)=\re \psi^*(x,t)\frac{\hbar}{im}\nabla\psi(x,t)
\eeq
\cite[(4.9)]{Mes1}. Thus an ensemble behaves like a flow of 
noninteracting particles.

\bigskip
{\bf Bohmian mechanics.}
Bohmian mechanics tries to give reality to this picture of an ensemble 
as a flow of particles with classical-like properties. 

Following {\sc Holland} \cite[Section 3.1]{Hol},
ensembles are interpreted in Bohmian mechanics as classical ensembles
of particles characterized by a solution $\psi(x,t)$ of Schr\"odinger's 
wave equation \gzit{e.schr}
and a trajectory $x(t)$ obtained by solving the initial value 
problem 
\lbeq{e.traj}
\dot x(t)=\frac{1}{m}\nabla S(x,t)\Big|_{x=x(t)},
\eeq
where the phase $S(x,t)$ of $\psi$ is defined by
\lbeq{e.phase}
\psi(x,t)=e^{iS(x,t)/\hbar}|\psi(x,t)|.
\eeq
The probability that a particle in the ensemble lies between the 
points $x$ and $x+dx$ at time $t$ is given by $|\psi(x,t)|^2dx$.
(Holland discusses the 3-dimensional case and hence has a volume 
element in place of $dx$. It would be trivial to rewrite the present 
discussion in three dimensions without changing the conclusion. 
Similarly, as in many expositions of Bohmian mechanics, spin is 
ignored, but incorporating it would not change anything essential.)

To indicate the flow of individual particles in an ensemble described
by a fixed solution $\psi(x,t)$ of the Schr\"odinger equation, we
refine the notation and write $x_\xi(t)$ for the position of a particle
that is in position $\xi$ at time $t=0$, so that $x_\xi(0)=\xi$.
The associated probability measure is then 
$d\mu(\xi)=|\psi_0(\xi)|^2d\xi$. Ensemble expectations of some 
real property $A_\xi$ that a particle -- characterized by its 
wave function $\psi_0$ (asumed fixed) and its position $\xi$ at time 
$t=0$ -- has are therefore given by averaging the values of $A$ over
the ensemble,
\lbeq{e.expB}
\<A\>_B=\int A_\xi|\psi_0(\xi)|^2d\xi.
\eeq
Since 
\lbeq{e.J}
J(x(t),t)=P(x(t),t)\dot x(t)
\eeq
\cite[(3.2.29)]{Hol},
the continuity equation \gzit{e.curr} implies that expectations 
of functions $A(x(t),t)$ are invariant under a shift of the 
reference time $t=0$. (Note that other authors use the equation 
\lbeq{e.traj2}
\dot x(t)=J(x(t),t)/P(x(t),t)
\eeq
in place of \gzit{e.traj} to define the trajectories; because of 
\gzit{e.J}, this is indeed equivalent and has the advantage of being 
directly motivated by time shift invariance.)

\bigskip
{\bf Local expectation values.}
To calculate expectation values of quantum mechanical operators, 
{\sc Holland} \cite[(3.5.4)]{Hol} defines the local expectation value 
of a Hermitian operator $A$ in the Schr\"odinger picture as the real 
number
\lbeq{e.loc}
A(x,t)=\re \frac{(A\psi)(x,t)}{\psi(x,t)}.
\eeq
The local expectation values evaluated along a trajectory, 
\lbeq{e.prop}
A_\xi(t)=A(x_\xi(t),t),
\eeq
are considered to be the real properties of a particle.
Indeed, Holland mentions in \cite[Section 3.7.2]{Hol} that the local 
expectation value ``might, following the common parlance, 
be termed the `hidden variable' associated with the corresponding 
physical variable''. With this definition of real properties, 
Bohmian mechanics achieves agreement with simple quantum mechanical 
predictions since, as is easily checked,
\lbeq{e.BeqQ}
\<A\>_B=\<A\>_Q
\eeq
({\sc Holland} \cite[(3.8.8/9)]{Hol}).
To appreciate what the local expectation values are in specific cases, 
Holland calculates explicitly the case of position, momentum, total 
energy, and total orbital angular momentum. In particular, the 
particle positions (local expectation values of $A=q$) and 
particle momenta (local expectation values of $A=p$) at arbitrary 
times $t$ are
\lbeq{e.qp}
q_\xi(t)=x_\xi(t),~~~p_\xi(t)=\nabla S(x_\xi(t),t)
\eeq
\cite[(3.2.18)]{Hol}. More generally, if $A=f(q)$ then $A(x,t)=f(x)$; 
thus functions of position at a fixed time behave classically. But for 
other operators, this is not the case; e.g., while 
$p_\xi(t)=m\dot x_\xi(t)$, the kinetic energy $K=p^2/2m$ satisfies 
\[
K_\xi(t)=\frac{m}{2}\dot x_\xi(t)^2+Q(x_\xi(t),t)
\]
with an additional `quantum potential' $Q(x,t)$.

\bigskip
\section{Time correlations in Bohmian mechanics} 

\bigskip
{\bf Particles in the ground state.}
For any Hamiltonian with a nondegenerate ground state $\psi_0$
(satisfying $H\psi_0=E_0\psi_0$), this ground state can always be taken
to be real. Indeed, since the complex conjugate $\psi_0^*$ also 
satisfies $H\psi_0^*=E_0\psi_0^*$ and the ground state is 
nondegenerate, $\psi_0^*$ must be a multiple of $\psi_0$, and scaling
with the square root of the multiplier leaves a real eigenfunction.

The solution $\psi$ of the Schr\"odinger equation \gzit{e.schr} 
corresponding to the ground state is
\[
\psi(x,t)=e^{-itE_0/\hbar}\psi_0(x).
\]
If a particle can be in position $x$ 
at time $t$ then $|\psi(x,t)|^2>0$, hence $\psi_0\neq 0$.
A comparison with \gzit{e.phase} therefore shows that particles in a 
nondegenerate ground state have a phase $S(x,t)=\pm tE_0$ independent 
of $x$. Thus \gzit{e.traj} implies that $x(t)$ is constant, 
$x_\xi(t)=\xi$ for all $t$. Thus each particle in the ensemble stands 
still. 

This observation is puzzling and lead Einstein to reject
the Bohmian interpretation; see {\sc Holland} \cite[Section 6.5.3]{Hol}
for a discussion and a defense.

\bigskip
{\bf The harmonic oscillator.}
A one-dimensional harmonic oscillator of mass $ m$, period $T$ and 
angular frequency $\omega=2\pi/T$ is described by the 
Hamiltonian
\lbeq{e1}
H(p,q)=\frac{p^2}{2m}+\frac{\omega^2m}{2}q^2.
\eeq
The
canonical commutation relations \gzit{e.ccr} imply that, for
the Hamiltonian \gzit{e1}, the Heisenberg dynamics \gzit{e.heis}
of position and momentum are given by
\[
\frac{dq(t)}{dt}=\frac{p(t)}{m},~~~\frac{dp(t)}{dt}=-\omega^2mq(t),
\]
just as in the classical case. In particular, we can solve the
dynamics explicitly in terms of the position operator $q$
and the momentum operator $p$ at time $t=0$ as
\[
q(t)=q\cos\omega t + \frac{p}{\omega m}\sin\omega t,
\]
\[
p(t)=p\cos\omega t - q\omega m\sin\omega t,
\]
again as in the classical case. In particular, $q(t+T/2)=-q(t)$, so 
that quantum mechanics predicts the time correlation
\lbeq{e.corQ}
\<q(t+T/2)q(t)\>_Q=-\<q(t)^2\>_Q <0
\eeq
for an ensemble in an arbitrary pure (or even mixed) state. 
($\<q(t)^2\>_Q=0$ would be possible only in an eigenstate of 
$q(t)$ to the eigenvalue zero, but there is no such normalized state.)

On the other hand, interpreting the time correlations in a Bohmian 
sense, one finds from \gzit{e.qp} and \gzit{e.expB} that
\[
\<q(t+T/2)q(t)\>_B=\int q_\xi(t+T/2)q_\xi(t)|\psi_0(\xi)|^2d\xi.
\]
For particles in the ground state (which for the harmonic oscillator 
is nondegenerate), the discussion above shows that the right hand
side is constant,
\lbeq{e.corB}
\<q(t+T/2)q(t)\>_B=\<q(t)^2\>_B=\<q(t)^2\>_Q > 0.
\eeq
Comparing \gzit{e.corQ} and \gzit{e.corB}, we see that the 
quantum mechanical time correlation and the Bohmian time
correlation have opposite signs.

\bigskip
{\bf Measuring time correlations.}
The fact that, in general, $q(s)q(t)$ is not Hermitian and hence 
cannot be measured in {\em individual} events does not mean that the 
expectation on the left hand side of \gzit{e.corQ} is meaningless and
has no relation to experiment. Indeed, one may
define the expectation of an arbitrary quantity $f$ in orthodox 
quantum mechanics (where all self-adjoint operators = observables
can be measured, cf. {\sc Dirac} \cite[p.37]{Dir}) 
in terms of the observables $\re f =\frac{1}{2} (f+f^*)$ and 
$\im f =\frac{1}{2i} (f-f^*)$  by 
\lbeq{e.meas}
\<f\>_Q :=\<\re f\>_Q +i\<\im f\>_Q.
\eeq
This gives unambiguous values to all expectations, and is fully 
consistent with orthodox quantum mechanics. Of course, it may 
not be easy to measure $\re f$ and $\im f$, but an operational 
procedure for measuring arbitrary Hermitian functions of $p$ and $q$ 
by a suitable experimental arrangement can be found, e.g., in 
{\sc Lamb} \cite{Lam}. And quantum optics routinely deals with
expectations and measurements of coherent states, which are eigenstates
of nonhermitian annihilator operators; see, e.g. {\sc Leonhardt}
\cite{Leo}.

While the example of the harmonic oscillator is somewhat artificial,
it has the advantage that all calculations can be done explicitly.
Significant physical applications of time correlations are, however,
made in statistical mechanics, where integrals over time correlations 
in thermodynamic equilibrium states are naturally linked to linear 
response functions, and hence are measurable as susceptibilities. 
See, e.g., {\sc Reichl} \cite[(15.161) and (15.172)]{Rei}. 
Time correlations also arise in the calculation of optical spectra
({\sc Carmichael} \cite[Lecture 3.3]{Car})
and in the context of quantum Markov processes 
({\sc Gardiner} \cite[Section 10.5]{Gar}).
Thus, at least in principle, it is possible to test the validity of
the recipe \gzit{e.meas} by experiment, by measuring susceptibilities
or spectra directly, and by comparing the result to that obtained by 
applying \gzit{e.meas} to measurements of $\re f$ and $\im f$.

As Arkadiusz Jadczyk (personal communication) pointed out,
\gzit{e.meas} implies that due to noncommutativity, the quantum 
mechanical time correlations $\<q(s)q(t)\>_Q$ are complex in most 
states at most times, while time correlations computed from Bohm 
trajectories are always real. Thus an agreement would be a coincidence.

On the other hand, it is possible to avoid nonhermitian operators 
completely. Indeed, the contradiction persists in the following 
consequence of \gzit{e.corQ} and \gzit{e.corB}:
\lbeq{e.corQh}
\<q(t+T/2)q(t)+q(t)q(t+T/2)\>_Q=-2\<q(t)^2\>_Q <0,
\eeq
\lbeq{e.corBh}
\<q(t+T/2)q(t)+q(t)q(t+T/2)\>_B=2\<q(t)^2\>_Q > 0.
\eeq
Note that $q(t+T/2)q(t)+q(t)q(t+T/2)$ is Hermitian, and 
\gzit{e.corQh} has the correct classical time correlation as limit 
when $\hbar \to 0$.
Symmetrized time correlations are discussed in the context of linear
response theory in {\sc Kubo} et al. \cite[pp. 167-169]{KubTH}.

\bigskip
In discussions with proponents of Bohmian mechanics, it is claimed that
my interpretation of the Bohmian formalism is erroneous, in that I
am not making the proper distinction between the ontological "beable" 
and the epistemological "observable", and compare the statistics of 
unobserved Bohm trajectories with those for quantum observations.

However, quantum mechanics can be used in practice without reference 
to the (still ill-defined) measurement mechanism, while Bohmian
mechanics resorts to the latter to justify any discrepancy. 
This should not be the case if the `beables' were the real entities 
that Bohmian mechanics claims them to be. And indeed, the whole 
purpose of the local expectation values is to show the equivalence
of expectations in Bohmian mechanics with those in quantum mechanics,
without having to refer to measurement. 

What else could the meaning of \gzit{e.BeqQ} be? The whole discussion 
in {\sc Holland} \cite[Section 3.5--3.8]{Hol} becomes meaningless unless
it is accepted that \gzit{e.BeqQ} is the real link between quantum
mechanics and Bohmian mechanics, independent of any measurement 
questions. The probabilities -- Holland discusses these 
independent of expectations -- follow the rule \gzit{e.BeqQ} when $A$ 
is an orthogonal projector corresponding to the associated subspaces,
and if the expectation rule fails then associated probabilities also 
fail. 

Thus, one wonders why Bohmian mechanics, which can do calculations
of single-time probabilities without reference to measurement 
questions, suddenly needs the measurement process to calculate 
probabilities of pair events occuring at two different times.

\bigskip
It may be noted that there are similar problems with multiple times in 
{\sc Nelson}'s \cite{Nel} stochastic quantum mechanics; 
{\sc Blanchard} et al. \cite{BlaGS} show how these problems can be 
overcome by a specific procedure for state reduction under measurement.

However, Bohmian mechanics does not seem to have such an option to 
rescue its interpretation since in the orthodox Bohmian 
interpretation, state reduction is a purely dynamical phenomenon.
The suggestion to explain equivalence to quantummechanical predictions 
by invoking the measurement process leads at best to an approximate
equivalence since Bohmian theory discusses measurement only in an 
approximate way ({\sc Holland} \cite[Chapter 8]{Hol}, 
{\sc Bohm \& Hiley} \cite[Chapter 6]{BohH}). And even then, specific
efforts would be needed to show that the time correlations come out
in the right way. 

And the explanation by measurement fails completely
if we consider the universe as a whole which, if supposed to behave
deterministically according to the laws of Bohmian mechanics,
has no meaningful way of defining time correlations apart from
$\<q(s)q(t)\>_B$.

\bigskip
{\bf The ambiguity of local expectation values.}
To gain a better understanding of the problems of Bohmian mechanics 
from a slightly different point of view, we look more 
closely at the local expectation values that are supposed to 
define the real properties of particles, and that lie at the heart of
the claim of Bohmian mechanics that all its predictions agree 
with those of quantum mechanics.

We first note that the recipe for calculating local expectation values
is linear without restriction; in particular, for a particle in the
ground state, where $q_\xi(t)=\xi$ and $p_\xi(t)=0$, we have 
$A_\xi(t)=\alpha \xi$ for any operator $A=\alpha q + \beta p$.
We use this to calculate the local expectation value of the Heisenberg 
position operator $A=q(s)$ at time $s$ in the ground state of the 
harmonic oscillator, and find the remarkable formula
\[
A_\xi(t)=\xi \cos \omega s.
\]
Thus the objective value of $A=q(s)$ at any time $t$ is 
$\xi \cos \omega s$, corresponding to our intuition if we regard $s$ as 
the physical time. It seems that, at least in the Bohmian picture 
of the harmonic oscillator, the Heisenberg time $s$ is the real time 
while the Schr\"odinger time $t$ only plays a formal and 
counterintuitive role. 

This gives weight to what is called `operator realism' in 
{\sc Daumer} et al. \cite{DauDG}, against the Bohmian program 
advocated there. And it makes the interpretation of
local expectation values as real properties of the system highly 
dubious since these values depend on the choice of the Heisenberg time 
$s$.

In particular, for multi-time expectations, which are meaningful
only in the Heisenberg picture, there is no distinguished single
Heisenberg time, and hence no natural Bohmian interpretation. 

Thus Bohmian mechanics can at best be said to reproduce a subset of 
quantum mechanics. It contradicts the quantum mechanical predictions 
about time correlations if one proceeds in the straightforward way
that generalizes the basic formula \gzit{e.BeqQ} that accounts for
agreement of single time expectations and single-time probabilities.

And Bohmian mechanics does not say anything at all about time 
correlations if the connection to quantum mechanics is kept more vague 
and left hidden behind a measurement process that is inherently 
approximate in Bohmian mechanics. Should this be the
real link between quantum mechanics and Bohmian mechanics, one 
could claim the predictions of Bohmian mechanics to be approximately 
equal onlyto those of quantum mechanics, against the explicit 
assertions of many supporters of Bohmian mechanics.

\bigskip
\section{Conclusion}

In contrast to the claim by {\sc D\"urr} et al. \cite{DueGZ},
Bohmian mechanics {\em does not} account for all of the phenomena 
governed by nonrelativistic quantum mechanics. Indeed, 
it was shown that for a harmonic oscillator in the ground state, 
Bohmian mechanics and quantum mechanics predict values of opposite 
sign for certain time correlations. Bohmian mechanics therefore 
contradicts quantum mechanics at the level of time correlations. 
Since time correlations can be observed experimentally via linear 
response theory, Bohmian mechanics and quantum mechanics cannot both 
describe experimental reality.

Due to the complicated form of the Bohmian dynamics, it seems 
difficult to compute time correlations for realistic scenarios where 
a comparison with linear response theory and hence with experiment 
would become possible. But perhaps numerical simulations are feasible.
On the other hand, it is unlikely that, if the predictions of quantum 
mechanics and Bohmian mechanics differ in such a simple case, they 
would agree in more realistic situations. 

The time correlations used in statistical mechanics are those from 
quantum mechanics and not those from Bohm trajectories. Moreover, they 
can be calculated and used without reference to any theory about 
the measurement process. If an elaborate theory of quantum observation 
is needed to reinterpret Bohmian mechanics -- so that it matches 
quantum mechanics and thus restores the connection to statistical 
mechanics -- then Bohmian mechanics is at best approximately 
equivalent to quantum mechanics and, I believe, irrelevant to practice. 

It is therefore likely that Bohmian mechanics is ruled out as a 
possible foundation of physics.

\bigskip

\end {document}